\documentclass[11pt]{article}

\usepackage[a4paper,hmargin=1.72cm,vmargin=2.4cm]{geometry}
\usepackage{amsmath,amssymb,amsfonts,mathtools}
\usepackage{bm}
\usepackage{bbm}
\usepackage{booktabs}
\usepackage{enumitem}
\usepackage{graphicx}

\usepackage{xcolor}
\usepackage{tikz}
\usetikzlibrary{arrows.meta,positioning,calc,shapes.geometric,fit}
\usepackage{caption}
\usepackage{subcaption}
\usepackage{hyperref}
\usepackage{microtype}
\usepackage[authoryear,round]{natbib}
\bibpunct[, ]{(}{)}{,}{a}{}{,}

\usepackage{authblk}
\hypersetup{
    colorlinks=true,
    linkcolor=blue!50!black,
    citecolor=blue!50!black,
    urlcolor=blue!50!black
}

\newcommand{\dd}{\mathrm{d}}
\newcommand{\calH}{\mathcal{H}}

\title{Hawkes-Driven OTC Market Making: \\
Volterra--Riccati Approximation}
\author{Alexander \textsc{Barzykin}\footnote{HSBC, 8 Canada Square, Canary Wharf, London E14 5HQ, United Kingdom, \texttt{alexander.barzykin@hsbc.com}.}}
\date{\today}

\begin{document}
\sloppy
\maketitle

\begin{abstract}
We formulate an over-the-counter (OTC) market-making problem in which request-for-quote (RFQ) arrivals are modelled by general Hawkes kernels and fills are controlled thinnings of the exogenous request flow. The modelling choice is motivated by spot-FX RFQ data: after filtering and transforming to seasonality-adjusted RFQ activity time, two-way activity in major currency pairs exhibits large fitted branching ratios and multi-scale persistence. For general Hawkes kernels the control problem is path-dependent: the relevant state contains the order-flow history, or equivalently the forward curve of conditional future RFQ intensities. Exact Markovian lifting is available for exponential kernels, but it becomes high-dimensional for mixtures of exponentials and impractical for long-memory kernels. We therefore develop a hierarchy of Volterra--Riccati approximations. The first level replaces random future request flow by its conditional Volterra forecast; the second adds a covariance correction for intensity uncertainty; the third updates the quote rule with the realized Hawkes memory, or equivalently with the post-request conditional forecast curve. Although Hawkes dynamics provide these inputs explicitly, the Riccati control layer uses only the conditional mean intensity, its covariance and the post-request update, and can therefore in principle be paired with other exogenous request-flow models. The approximation hierarchy is validated in an exponential Hawkes benchmark, where the exact lifted HJB can be solved numerically. The state-feedback Volterra--Riccati policy closely tracks the exact benchmark, especially in directional regimes, while a memory-free Poisson policy suffers substantial regret. We then apply the same state-feedback rule to a power-law-like RFQ memory model. A directional RFQ burst changes the conditional forecast of future flow and is converted by the continuation-value shadow price into a persistent quote skew. The resulting endogenous OTC quote impact inherits the long-memory decay of the RFQ forecast response and improves inventory and P\&L risk control relative to a no-conditioning Poisson benchmark.
\end{abstract}

\vspace{7mm}


\vspace{5mm}

\section{Introduction}

Classical stochastic-control models of market making represent buy and sell arrivals as controlled point processes whose intensities decrease with the quoted distance from the mid-price. The dealer earns spread revenue while managing running and terminal inventory risk. This approach originates in dealer models such as Ho--Stoll \citeyearpar{HoStoll1981} and Avellaneda--Stoikov \citeyearpar{AvellanedaStoikov2008} and has been further developed into the Hamiltonian framework \citep{GueantLehalleFernandezTapia2013,Gueant2016,CarteaJaimungalPenalva2015}. \citet{BergaultEvangelistaGueantVieira2021} showed that quadratic expansions of the OTC Hamiltonians lead to Riccati equations and tractable quote decompositions in high-dimensional market-making problems. \citet{BergaultGueant2023} further extend micro-price ideas to OTC RFQ markets and model liquidity imbalances through bi-dimensional Markov-modulated Poisson processes, providing a pricing framework for illiquid or one-sided RFQ markets. Their distinction between an underlying liquidity state and the dealer's price-dependent response is related in spirit to the request/response separation used below.

The Poisson specification is analytically convenient but cannot represent clustered and persistent order flow. Hawkes processes are a flexible reduced form alternative and are widely used in market microstructure to represent clustering, order-flow persistence and liquidity resilience \citep{Large2007,BacryMastromatteoMuzy2015,HardimanBercotBouchaud2013,LallouacheChallet2016}. Nearly unstable Hawkes processes provide one mechanism for long-memory limits \citep{JaissonRosenbaum2015}. Persistent signed flow is also central to theories of market impact and volatility: market impact can be interpreted as anticipation of future order-flow imbalance \citep{Jaisson2015Impact}, no-arbitrage restrictions lead to power-law impact and rough volatility \citep{JusselinRosenbaum2020}, and recent unified models connect order-flow persistence, impact and volatility through Hawkes reaction flows \citep{MuhleKarbeChahdiRosenbaumSzymanski2026}. Importantly for the OTC motivation, \citet{EislerBouchaud2016} study credit-index transactions without a limit order book and find persistent order-flow correlations and impact behaviour compatible with more classical electronic markets.

We do not attach a structural contagion interpretation to the Hawkes branching representation. Correlated RFQs may be reactions to a common unobserved cause---public market activity, volatility, a client execution programme or another latent liquidity factor---rather than consequences of earlier RFQs. If such exogenous variation is omitted from the baseline intensity, a fitted Hawkes model may absorb part of it into the excitation kernel and thereby produce a large branching ratio. In this paper the Hawkes specification is therefore used primarily as a predictive model for the conditional distribution of future RFQ flow. The control problem depends on that conditional forecast, not on a causal interpretation of its statistical decomposition.

This reduced-form interpretation is consistent with a broader effective-memory viewpoint in stochastic dynamics. In generalized Langevin models, eliminating latent degrees of freedom can leave an observable process with a memory kernel and stochastic forcing, so persistent memory need not be read as direct interaction between successive observations. Recent financial applications separate the memory kernel, nonlinear potential and noise covariance \citep{Itkin2026BeyondRough}, while the Lean Marketron model provides an example in which the log-price admits a state-modulated generalized Langevin representation and the memory variable gives an exact Markovian lift of that representation \citep{Itkin2026LeanMarketron}.

The closest market-making reference is \citet{Jusselin2021}, who considers general Hawkes-driven market-order flow in an electronic market, formulates the resulting non-Markovian stochastic-control problem, proves existence and uniqueness of a viscosity solution to the associated HJB equation, and develops a consistent numerical approximation. The common starting point is that the relevant state is the full order-flow history, or an equivalent forward intensity curve. The present paper differs in two respects. First, it is built around the OTC distinction between an exogenous RFQ opportunity and its controlled conversion into a fill: observing a request changes the information state even when the dealer does not trade. Second, rather than approximating the full path-dependent value function, we seek a low-dimensional and interpretable quote rule by combining the conditional Hawkes forecast with the quadratic/Riccati approximation of \citet{BergaultEvangelistaGueantVieira2021}. The exact exponential lift is used as a validation benchmark, while the intended application is to general long-memory kernels.

Other uses of Hawkes dynamics in execution and market making include execution with mixed market-impact Hawkes price dynamics \citep{AlfonsiBlanc2016}, linear-quadratic Volterra control \citep{AbiJaberMillerPham2021}, liquidation with general Volterra propagator impact and signals \citep{AbiJaberNeuman2025}, mutually exciting market-order and book-shape models \citep{CarteaJaimungalRicci2018}, and numerical Hawkes limit-order-book solvers based on deep-learning or impulse-control methods \citep{Kumar2024,JainFiroozyeKochemsTreleaven2025}. \citet{MartinTse2026} use a Markovian approximation of Hawkes kernels via Mercer's expansion and solve the resulting high-dimensional HJB with a PINN. \citet{BergaultBertucciBoubaGueantGuilbert2024} discuss related Hawkes liquidity states in price-aware automated market makers and emphasize that when order-flow states jump together with inventory, exact dynamic programming becomes high-dimensional.

More broadly, the paper is connected to the literature on non-Markovian and path-dependent stochastic control. \citet{BandiniKeller2024} study stochastic control of path-dependent piecewise deterministic processes and the associated non-local path-dependent Hamilton--Jacobi--Bellman equations. \citet{AllanCohen2020} develop a pathwise stochastic-control approach based on rough differential equations, while \citet{BankEtAl2025} use path signatures to parameterise progressively measurable controls for generic non-Markovian stochastic-control problems. These works provide general path-dependent control frameworks. The approach below is more specialized: it exploits the Hawkes--Volterra structure of RFQ forecasts and the quadratic expansion of market-making Hamiltonians to obtain an interpretable low-dimensional quote approximation. Accordingly, we treat the conditional forecast curve as the economically relevant reduced state; when a finite-dimensional Markovian lift exists, it is an alternative computational representation rather than a prerequisite for the control construction.

A useful precedent comes from non-Markovian optimal execution \citep{AbiJaberNeuman2025}. In linear-quadratic execution models with transient impact, one can often avoid a high-dimensional Markovian lift: convexity and the linear-quadratic structure lead, through the stochastic maximum principle or equivalent first-order conditions, to Fredholm, Volterra or Riccati--Volterra equations for the optimal trading rate. Memory therefore does not by itself preclude tractable control. The present market-making problem is less direct. The dealer controls quote-dependent conversion probabilities, the Hamiltonians are nonlinear, and RFQ arrivals change the future distribution of inventory. The method developed below follows the same non-Markovian philosophy---work with conditional forecast curves rather than a large lifted state---but adapts it to nonlinear quote-driven market making.

We consider an OTC/RFQ market in which requests are exogenous information events and executions are quote-dependent conversions of those requests. The dealer's quote changes the probability of winning a request but does not create the request itself. Economically, the same two stages are present in a request-for-stream setting even when the decomposition is not separately recorded: there is first an intention or willingness to trade and then a decision to trade conditional on the prices received. This separation makes the conditional forecast of future request flow a Volterra object driven by observed RFQ history, while the dealer's conversion decision remains a market-making control.

We begin by documenting persistent RFQ activity in spot-FX data after filtering and transformation to seasonality-adjusted RFQ activity time. We then formulate the exact path-dependent dynamic program for RFQ market making with general Hawkes kernels, separating the information value of observing a request from the execution value of winning it. We derive a hierarchy of Volterra--Riccati approximations: a conditional-mean policy, a noise-aware covariance correction, and a state-feedback rule that updates the quote with the post-request forecast curve. Finally, we validate the hierarchy in an exponential Hawkes benchmark where the exact lifted HJB is available, and apply the state-feedback rule to a power-law-like RFQ memory model. In the long-memory experiment, directional RFQ bursts generate persistent quote skew through the continuation-value shadow price, giving an endogenous OTC analogue of market impact and reducing inventory and P\&L risk relative to a memory-free Poisson benchmark.

\section{Empirical motivation: persistent RFQ activity}
\label{sec:empirical-motivation}

Before introducing the theoretical model, we briefly document the empirical feature that motivates it. We use anonymised spot-FX RFQ arrivals for EURUSD, GBPUSD and USDJPY.\footnote{HSBC data.} Since requests are typically two-way, the empirical process studied here is limited to an activity process rather than a signed buy/sell process. The raw data are filtered to retain ordinary trading days.  Weekends, major holidays, daylight-saving transition periods, the Christmas--New Year period, and daily roll windows with abnormal liquidity are removed.  The active window is taken to be $[00{:}00,20{:}00]$ in London time.  We also remove very small requests and client streams with negligible historical conversion, in order to reduce price-discovery traffic with little apparent trading intent.  In the results below we keep RFQs of size at least $0.25$ million notional and clients with historical fill ratio of at least 1\%. RFQ sizes are assigned to a small number of standard FX size buckets. The historical window spans three years and the number of qualified events after filtering is in excess of $5\cdot 10^5$ for each currency pair.

FX RFQ activity has strong deterministic intraday seasonality.  Estimating a Hawkes kernel directly in calendar time would therefore confound deterministic seasonality with residual serial dependence.  We consequently estimate the kernel in seasonality-adjusted RFQ activity time.  For each currency pair $c$, let $\bar\lambda^c(\theta)$ denote the average intraday RFQ activity profile at time of day $\theta$, normalised to have unit average over the active window.  Calendar time $t$ is then transformed to activity time
\begin{equation}
\tau^c(t) = \int_0^t \bar\lambda^c(\theta(u))\,\dd u .
\label{eq:empirical-business-time}
\end{equation}
The terminology \emph{activity time} is deliberate: unlike event-count or transaction-time clocks that are also sometimes called business time, the transformation \eqref{eq:empirical-business-time} is deterministic. The fitted kernel should therefore be interpreted as an activity-time predictive memory kernel rather than a wall-clock causal response.

All rates entering a control problem must be expressed in the same clock. If $u=\tau^c(t)$, $t^c(u)=(\tau^c)^{-1}(u)$ and the calendar-time variance rate is $(\sigma_{\rm cal}^c(t))^2$, then the time-changed price has variance rate
\[
    \bigl(\widetilde\sigma^c(u)\bigr)^2
    = \bigl(\sigma_{\rm cal}^c(t^c(u))\bigr)^2\frac{\dd t^c(u)}{\dd u}.
\]
Writing $q_u:=q_{t^c(u)}$, consequently,
\[
    \frac{\gamma}{2}\int \bigl(\sigma_{\rm cal}^c(t)\bigr)^2q_t^2\,\dd t
    =\frac{\gamma}{2}\int \bigl(\widetilde\sigma^c(u)\bigr)^2q_u^2\,\dd u.
\]
The present empirical exercise does not estimate $\widetilde\sigma^c$ and does not test whether volatility is constant in activity time. In the theoretical model below, time denotes one chosen clock and the volatility coefficient is understood per unit of that clock. Taking it constant is a separate modelling approximation used in the numerical examples, not a consequence of the activity-time transformation.

For each currency pair $c$ and size bucket $k$, let $M^{c,k}$ be the two-way RFQ counting process in activity time.  We fit the reduced form marked Hawkes activity model
\begin{equation}
\lambda_\tau^{c} = \mu^c  + \sum_k \alpha^{c,k} \int_0^\tau \varphi^c(\tau-u)\,\dd M_u^{c,k},
\label{eq:empirical-size-bucket-hawkes}
\end{equation}
where the size-bucket coefficient $\alpha^{c,k}$ is the total fitted excitation mass associated with an RFQ in bucket $k$.  The common unit-mass kernel is represented as a short mixture of exponentials,
\begin{equation}
\varphi^c(u) = \sum_n w_n^c\beta_n e^{-\beta_n u}, \qquad \sum_n w_n^c=1, \qquad \beta_n=\frac{\log 2}{h_n},
\label{eq:empirical-exp-mixture}
\end{equation}
with half-lives $h_n=1$, $10$ and $60$ minutes in activity-time units in this study.  If $\pi^{c,k}$ denotes the empirical fraction of RFQs in bucket $k$, the average branching ratio is
\begin{equation}
\eta^c = \sum_k \pi^{c,k} \alpha^{c,k}.
\label{eq:empirical-branching-ratio}
\end{equation}
Within the fitted branching representation, it is the average total offspring mass assigned to one additional request; it should not be interpreted as direct evidence that one client's RFQ causes another client's RFQ.

\begin{table}[t]
\centering
\caption{Reduced form Hawkes estimates for two-way RFQ activity in activity time.  The kernel is a three-exponential mixture with half-lives of $1$, $10$ and $60$ minutes.  The branching ratio $\eta$ is the size-weighted excitation mass in \eqref{eq:empirical-branching-ratio}.}
\label{tab:empirical-hawkes}
\begin{tabular}{lrrrr}
\toprule
Pair & $\eta$ & $w_{1}$ & $w_{10}$ & $w_{60}$ \\
\midrule
USDJPY & $0.862$ & $0.559$ & $0.320$ & $0.120$ \\
GBPUSD & $0.895$ & $0.521$ & $0.359$ & $0.120$ \\
EURUSD & $0.880$ & $0.692$ & $0.209$ & $0.099$ \\
\bottomrule
\end{tabular}
\end{table}

\begin{figure}[h]
\centering
\includegraphics[width=0.62\linewidth]{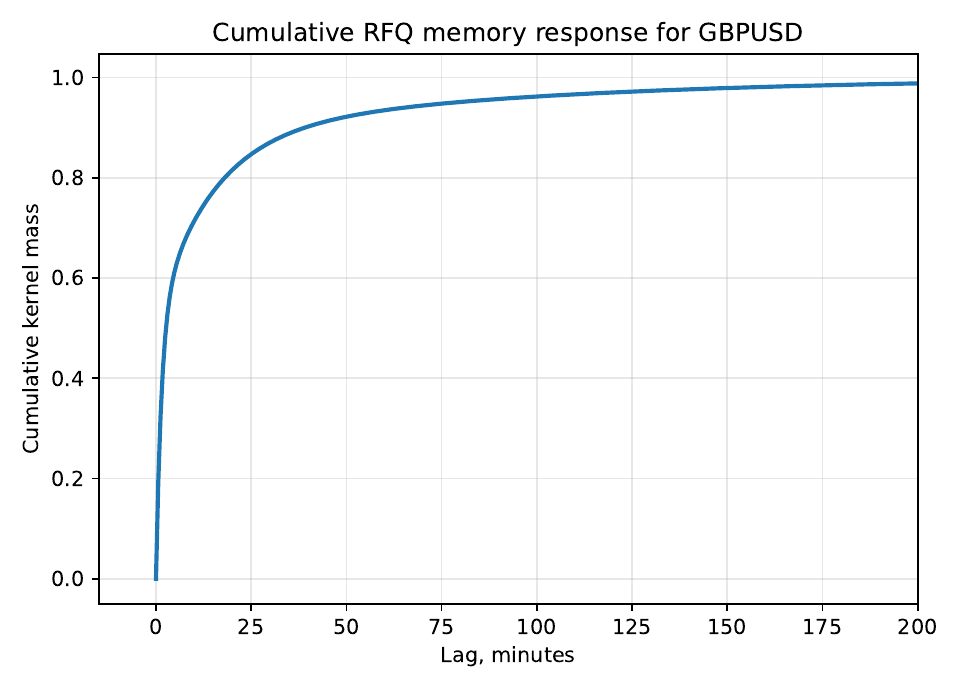}
\caption{Cumulative fitted RFQ memory response for GBPUSD.  The kernel was estimated from two-way RFQ request arrivals after filtering and transforming to activity time, as described in the text.}
\label{fig:empirical-cumulative-kernel}
\end{figure}

Table \ref{tab:empirical-hawkes} reports the fitted branching ratios and mixture weights.  Across all three major pairs, the fitted branching ratio is consistently high, around $0.86$--$0.90$, even after filtering and activity-time transformation.  The estimated kernels are multi-scale: a large part of the response is concentrated at minute horizons, but a non-negligible component persists over tens of minutes.
Figure \ref{fig:empirical-cumulative-kernel} plots the cumulative memory response
\begin{equation}
F^c(u) = \int_0^u \varphi^c(v)\,\dd v
\label{eq:empirical-cumulative-kernel}
\end{equation}
for GBPUSD. The RFQ response indeed does not vanish immediately after a request burst -- a material fraction persists over economically relevant intraday horizons.  Thus, for a dealer choosing quotes over a finite horizon, the conditional forecast of future RFQ activity is a relevant state variable.

The estimates in Table \ref{tab:empirical-hawkes} are used only as reduced form motivation. They are not intended to identify the structural source of RFQ bursts. A burst may reflect public market activity, correlated client demand, repeated price discovery, execution slicing, information propagation across dealers and venues, or a common latent factor omitted from the baseline intensity. In the latter case, part of the fitted branching mass may represent unmodelled exogenous variation rather than endogenous propagation. For the market-making control problem, these mechanisms enter through the same object: the dealer's conditional forecast of future RFQ activity. The empirical evidence therefore supports replacing a memoryless Poisson forecast by a history-dependent Hawkes/Volterra forecast, without requiring a causal interpretation of the fitted kernel.

The empirical model merges bid and ask sides because the observed requests are predominantly two-way. Section~\ref{sec:model} reintroduces side marks because the inventory consequence of a fill is directional. For one-way RFQs the side may be observed; for two-way RFQs it must be interpreted as a latent trade-intention mark or estimated through a filtering layer. The theoretical side-marked model is therefore an idealized control model rather than a claim that signed request flow is directly observed in the data used above.

\section{Hawkes-driven OTC market-making problem}
\label{sec:model}

\subsection{Request flow and dealer objective}

Let $T>0$ be a finite horizon and let $(\Omega,\mathcal{F},(\mathcal{F}_t)_{0\le t\le T},\mathbb{P})$ satisfy the usual conditions. Time $t$ is measured in one fixed clock. If the activity-time clock of Section~\ref{sec:empirical-motivation} is used, the volatility and all intensity coefficients are transformed to that clock as described there. For notational simplicity, we assume a constant variance rate in the chosen clock and write
\begin{equation}
    \dd S_t = \sigma\,\dd W_t,
\end{equation}
where the Wiener process $W$ is independent of the innovations driving the request process. Let $\mathcal{Z}=\{z_1,\ldots,z_K\}\subset\mathbb{R}_+$ be a discrete RFQ size ladder.  
Side $s = b$ (bid) denotes a client selling to the dealer, so a fill increases dealer inventory. 
Similarly, side $s = a$ (ask) denotes a client buying from the dealer, so a fill decreases dealer inventory. We define dealer directionality indicator $\epsilon_s$ for $s \in \{b,a\}$, where $\epsilon_b = + 1$, $\epsilon_a = - 1$.

The exogenous RFQ processes are denoted by $M^{s,k}$, where $k$ indexes size $z_k$. Their intensities are Hawkes processes with general kernels,
\begin{equation}
    \lambda_t^{s,k} = \mu^{s,k}(t) + \sum_{r\in\{b,a\}}\sum_{\ell=1}^K \int_0^t \varphi_{s,k;r,\ell}(t-u)\,\dd M_u^{r,\ell},
    \label{eq:general-hawkes}
\end{equation}
where $\mu^{s,k}(t)$ is a deterministic, locally bounded background intensity for side $s$ and size $z_k$; it is constant in the stationary numerical examples. A stochastic background intensity would be an additional state variable and is not included in the present model. The kernels are non-negative and locally integrable, and may be exponential, sums of exponentials, power-law-like, cross-exciting or side-asymmetric. In the stationary case, stability requires the spectral radius of the matrix of $L^1$ kernel masses to be less than one.

The important modelling convention is that \emph{requests}, not the dealer's controlled executions, drive the Hawkes memory. In an RFQ market the arrival of a request is an information event: it may reflect client demand, dealer-to-dealer dissemination, aggregator activity, broader market conditions or a common latent factor. The dealer's quote controls whether that request is converted into a trade, but a marginally more aggressive quote should not, in the base model, create the client's request itself. The same economic decomposition applies to request-for-stream protocols even when intention and conversion are not separately recorded. This convention also prevents the dealer from mechanically controlling future RFQ intensity through their own fill decisions. It can be relaxed when modelling public trade prints, last-look rejection feedback, client relationship scores, or hedging/externalisation flow. Those effects are deliberately separated from the baseline request-flow model.

Conditional on an RFQ of side $s$ and size $z_k$, the dealer posts offset $\delta_t^{s,k}$ and wins the request with probability
\begin{equation}
    f_k(\delta_t^{s,k}), \qquad f_k'(\delta)<0, \qquad 0<f_k(\delta)<1.
\end{equation}
For the analytical Hamiltonian on $\mathbb R$, we impose the classical regularity conditions $f_k\in C^2(\mathbb R;(0,1))$, $f_k'<0$,
\begin{equation}
    \lim_{\delta\to+\infty}\delta f_k(\delta)=0,
    \qquad
    \sup_{\delta\in\mathbb R}
    \frac{f_k(\delta)f_k''(\delta)}{(f_k'(\delta))^2}<2.
    \label{eq:classical-f-condition}
\end{equation}
These conditions make the first-order map strictly monotone and give a finite Hamiltonian with a unique optimizer. In the numerical schemes, quotes are additionally restricted to a compact interval $\mathcal A=[\underline\delta,\overline\delta]$; existence is then automatic, and every formula below applies with the supremum taken over $\mathcal A$. Exponential response functions remain useful for closed-form checks and can be handled on this compact domain. For RFQ markets a sigmoid win probability is often natural.\footnote{Here we consider side-symmetric win probabilities for simplicity; generalisation is straightforward.} The controlled execution processes $N^{s,k}$ are thinnings of the request processes and have predictable intensities
\begin{equation}
    \Lambda_t^{s,k}(\delta_t^{s,k}) = \lambda_t^{s,k} f_k(\delta_t^{s,k}).
\end{equation}
Inventory evolves as
\begin{equation}
    \dd q_t = \sum_{s\in\{b,a\}}\sum_{k=1}^K \epsilon_s z_k\,\dd N_t^{s,k}.
    \label{eq:inventory-dynamics}
\end{equation}

Let $\mathcal X_t$ denote the dealer's cash account.  With the above sign convention,
\begin{equation}
    \dd \mathcal X_t = - \sum_{s,k} \epsilon_s z_k(S_t-\epsilon_s \delta_t^{s,k})\,\dd N_t^{s,k}.
\end{equation}
Consequently
\begin{equation}
    \dd(\mathcal X_t+q_t S_t) = q_{t-}\,\dd S_t + \sum_{s,k}z_k\delta_t^{s,k}\,\dd N_t^{s,k}.
    \label{eq:wealth-dynamics}
\end{equation}
The inventory process is càdlàg. We denote by $q_{t-}$ its predictable
left limit, which is the inventory observed when quotes are chosen and
immediately before an execution jump. In ordinary time integrals we write
$q_t$ for simplicity, since $q_t$ and $q_{t-}$ differ only at jump times.

The dealer wants to maximize the expected marked-to-market value at the end of the trading period, $\mathcal X_T+q_T S_T$, while managing the risk associated with holding the inventory. Under the martingale mid-price assumption, the control-relevant reduced objective is given by \citet{CarteaJaimungalPenalva2015,Gueant2016}
\begin{equation}
    \sup_{\delta}\,\mathbb{E}\left[ \int_0^T \sum_{s,k} z_k\delta_t^{s,k}\,\dd N_t^{s,k} - \frac 12 \gamma \int_0^T \sigma^2 q_t^2\,\dd t -\ell(q_T) \right],
    \label{eq:objective}
\end{equation}
where $\gamma$ is the inventory risk aversion coefficient and $\ell(q_T)$ is the terminal penalty.

\subsection{Path-dependent dynamic program and limits of exact lifting}

For $t\in[0,T]$, let $\mathsf{H}_t$ denote the set of finite marked RFQ histories before time $t$,
\[
    H=\bigl((t_j,s_j,k_j)\bigr)_{j=1}^n,
    \qquad 0<t_1<\cdots<t_n<t,
\]
including the empty history. For a fixed history $H$, the associated counting measures can be written as
\[
    \dd M_u^{r,\ell}(H)
    =
    \sum_{j=1}^n
    \mathbbm{1}_{\{s_j=r,\,k_j=\ell\}}
    \,\delta_{t_j}(\dd u),
\]
so that
\[
    \int_0^t \varphi_{s,k;r,\ell}(t-u)\,\dd M_u^{r,\ell}(H)
    =
    \sum_{j=1}^n
    \mathbbm{1}_{\{s_j=r,\,k_j=\ell\}}
    \varphi_{s,k;r,\ell}(t-t_j).
\]
Thus the stochastic-integral representation in \eqref{eq:general-hawkes} becomes, on a fixed history, the history-dependent intensity functional
\begin{equation}
    \lambda^{s,k}(t,H)
    =\mu^{s,k}(t)
    +\sum_{j=1}^n \varphi_{s,k;s_j,k_j}(t-t_j).
    \label{eq:history-intensity-functional}
\end{equation}
If $\calH_t$ is the realized RFQ history, then the predictable stochastic intensity in \eqref{eq:general-hawkes} is simply
\[
    \lambda_t^{s,k}=\lambda^{s,k}(t,\calH_{t-}).
\]
For a new request of type $(s,k)$ at time $t$, write
\[
    H\oplus(t,s,k)
\]
for the concatenated history. The state variable of the exact control problem is therefore $(t,q,H)$ and the value function is the deterministic functional $V(t,q,H)$ on the history state space.

Define the per-size, per-request Hamiltonian
\begin{equation}
    h_k(p)=\sup_{\delta\in\mathbb{R}} f_k(\delta)(\delta-p),
    \label{eq:bg-hamiltonian}
\end{equation}
with unique optimizer $\delta_k^*(p)$ under the conditions above. If compact quote controls are used, the same notation denotes the supremum over $\mathcal A$. The post-request shadow price is
\begin{equation}
    p_k^s(t,q,H) =
    \frac{V(t,q,H\oplus(t,s,k))-V(t,q+\epsilon_s z_k,H\oplus(t,s,k))}{z_k}.
    \label{eq:path-shadow-price}
\end{equation}
Formally, the path-dependent HJB is the deterministic equation
\begin{equation}
\begin{aligned}
    0={}&\partial_t V(t,q,H)-\kappa q^2 \\
    &+\sum_{s,k}\lambda^{s,k}(t,H)
    \Big[V(t,q,H\oplus(t,s,k))-V(t,q,H)
    +z_k h_k\!\left(p_k^s(t,q,H)\right)\Big],
\end{aligned}
\label{eq:path-dependent-hjb}
\end{equation}
with $\kappa:=\frac12\gamma\sigma^2$ and terminal condition $V(T,q,H)=-\ell(q)$. Here $\partial_t$ denotes the horizontal derivative obtained by advancing time while appending no new RFQ. The first difference inside the brackets is the value effect of observing a request even if it is not won; the Hamiltonian is the incremental value of optimally converting that request into a trade. Equation \eqref{eq:path-dependent-hjb} is a deterministic equation on the history state space; stochasticity enters when it is evaluated along the realized state $(q_t,\calH_t)$.

This history-state formulation is closest to the exact general Hawkes treatment of \citet{Jusselin2021}. The additional OTC feature is that the history is updated by the request itself, whereas inventory is updated only by the controlled Bernoulli conversion. The approximation developed below does not replace Jusselin's path-dependent characterization; it targets a different computational object, namely a low-dimensional quote approximation expressed through conditional forecast curves.

Equation \eqref{eq:path-dependent-hjb} is the exact control problem, but it is not computationally viable for general kernels.  Exponential kernels yield finite-dimensional Markovian lifts, and sums of exponentials yield one state variable per factor, side and size.  Thus a $d$-factor approximation of a two-sided, $K$-size Hawkes model already introduces $2Kd$ flow-memory variables in addition to inventory.  A power-law kernel requires either a large exponential mixture or a genuinely infinite-dimensional Volterra state.

This obstruction is not simply the presence of memory.  In linear-quadratic execution, the stochastic maximum principle or equivalent first-order optimality conditions often reduce non-Markovian transient-impact problems to deterministic Volterra or Fredholm equations.  The RFQ market-making problem is different because the control acts through nonlinear quote-response functions and controlled jump intensities; after an RFQ, the value of a possible inventory jump has to be evaluated under an updated Hawkes forecast.  A direct maximum-principle route therefore does not yield the same closed finite-dimensional structure.

A Markovian lift combined with a quadratic approximation in inventory does not remove this difficulty because the expansion coefficients remain functions of the entire lifted Hawkes state.  Moreover, RFQ arrivals jump both the inventory and the memory state, producing nonlocal terms in the coefficient equations.  One therefore obtains a system of transport-jump equations in the memory variables, not a small Riccati system.  Machine-learning methods face the same structural issue: they may approximate a particular high-dimensional lifted benchmark, but they do not by themselves deliver a transparent low-dimensional quote formula, nor do they scale naturally to long-memory kernels without a large Markovian embedding.  The purpose of the Volterra--Riccati approximation below is to preserve the economically relevant conditional Hawkes forecast while avoiding the full lifted dynamic program.

\section{Volterra--Riccati approximation hierarchy}

\subsection{Conditional Volterra forecast}

For each current history $\calH_t$, define the conditional future intensity curve
\begin{equation}
    m_t^{s,k}(u)=\mathbb{E}\left[\lambda_{t+u}^{s,k}\mid\mathcal{F}_t\right],
    \qquad u\ge0.
\end{equation}
In vector form, with $m_t(u)$ collecting all side-size components, the Hawkes structure gives the deterministic Volterra equation
\begin{equation}
    m_t(u)=\zeta_t(u)+\int_0^u \Phi(u-v)m_t(v)\,\dd v,
    \label{eq:volterra-forecast}
\end{equation}
where $\Phi$ is the matrix kernel and
\begin{equation}
    \zeta_t(u)=\mu(t+u)+\int_0^t \Phi(t+u-r)\,\dd M_r
\end{equation}
is the contribution of the observed past.  This forecast is independent of the dealer's future quotes because quotes thin requests into executions but do not create requests.

It is useful to introduce the forecast response to a new request of type $i=(s,k)$.  Let $e_i$ be the unit vector corresponding to that type.  The additional conditional mean generated by observing such a request at the current time is the resolvent response $\rho_i$, defined by
\begin{equation}
    \rho_i(u)=\Phi(u)e_i+\int_0^u \Phi(u-v)\rho_i(v)\,\dd v.
    \label{eq:forecast-response}
\end{equation}
Thus, immediately after observing a type-$i$ request, the forward forecast curve changes from $m_t$ to $m_t+\rho_i$.

\subsection{Mean Volterra--Riccati approximation}

At a decision time $t$, the mean Volterra approximation freezes the conditional forecast curve $m_t$ and replaces the random future RFQ intensity by this deterministic curve. For $\tau\in[t,T]$, define the conditional mean approximation $v_0(\tau,q;m_t)$ by
\begin{equation}
    0 =\partial_\tau v_0(\tau,q)-\kappa q^2 + \sum_{s,k} m_t^{s,k} (\tau - t) z_k h_k\!\left( \Delta_k^s v_0(\tau,q) \right),
\label{eq:mean-vr-hjb}
\end{equation}
with terminal condition $v_0(T,q;m_t)=-\ell(q)$, where
\begin{equation}
    \Delta_k^s v(\tau,q)=\frac{v(\tau,q)-v(\tau,q+\epsilon_s z_k)}{z_k}.
\end{equation}
Equation~\eqref{eq:mean-vr-hjb} is not claimed to be the HJB equation of the original Hawkes control problem, nor a time-consistent auxiliary value function as the conditioning time $t$ changes. It is a certainty-equivalent approximation to the continuation value conditional on the information available at time $t$. Once $m_t$ is supplied, neither the Hawkes kernel $\Phi$ nor the request process $M$ appears in \eqref{eq:mean-vr-hjb}; Hawkes enters this level only through the forecast-generation step \eqref{eq:volterra-forecast}. Operationally, the forecast curve and backward Riccati system are recomputed as new information arrives, yielding a receding-horizon feedback policy. Section~\ref{sec:noise-aware} adds the leading covariance correction around this conditional mean.
The BEGV approximation \citep{BergaultEvangelistaGueantVieira2021} expands $h_k$ locally in the shadow price,
\begin{equation}
    h_k(p)\approx h_{k,0}+h_{k,1}p+\frac12 h_{k,2}p^2,
    \label{eq:h-expansion}
\end{equation}
and uses a quadratic inventory ansatz
\begin{equation}
    v_0(\tau,q;m_t)\approx -\frac12 A_0(\tau)q^2 + B_0(\tau)q + C_0(\tau).
    \label{eq:quadratic-ansatz}
\end{equation}
For this ansatz,
\begin{equation}
    \Delta_k^s v_0(\tau,q) = \epsilon_s A_0(\tau)q - \epsilon_s B_0(\tau) + \frac12 A_0(\tau) z_k.
    \label{eq:quadratic-shadow}
\end{equation}
Substitution of \eqref{eq:h-expansion} and \eqref{eq:quadratic-shadow} into \eqref{eq:mean-vr-hjb}, followed by coefficient matching in $q$, gives a backward Riccati system. 
Writing
\begin{equation}
    d_{s,k}(\tau)=-\epsilon_s B_0(\tau)+\frac12 A_0(\tau)z_k,
\end{equation}
the coefficient equations take the representative form
\begin{align}
    \dot A_0(\tau) &= -2\kappa + A_0(\tau)^2\sum_{s,k} m_t^{s, k}(\tau-t) z_k h_{k,2},
    \label{eq:A0-riccati}\\
    \dot B_0(\tau) &= -A_0(\tau) \sum_{s, k}\epsilon_s m_t^{s, k}(\tau-t) z_k \left( h_{k, 1} + h_{k, 2}d_{s, k}(\tau) \right),
    \label{eq:B0-riccati}\\
    \dot C_0(\tau) &= -\sum_{s, k} m_t^{s, k}(\tau-t) z_k \left( h_{k, 0} + h_{k, 1} d_{s, k}(\tau) + \frac12 h_{k, 2} d_{s, k}(\tau)^2 \right),
    \label{eq:C0-riccati}
\end{align}
with terminal values determined by $\ell(q)$.\footnote{The coefficient equations close cleanly when the terminal penalty is quadratic, e.g. $\ell(q) = \kappa_T q^2$, with terminal conditions $A_0(T) = 2\kappa_T$, $B_0(T) = C_0(T) = 0$. If $\ell(q)$ is general then the quadratic ansatz requires a terminal quadratic projection.}  For symmetric side intensities and penalties the skew coefficient $B_0$ is zero; for directional bursts it becomes the dominant quote-skew term.  At each decision time, after updating the conditional forecast and resolving the approximation, the quote itself is computed from the exact Hamiltonian optimizer,
\begin{equation}
    \delta_t^{s,k}=\delta_k^*\!\left(\Delta_k^s v_0(t,q_t;m_t)\right),
    \label{eq:mean-vr-quote}
\end{equation}
so the Hamiltonian expansion is used to approximate the continuation value, not to impose a linear quote rule.

\subsection{Noise-aware second-order correction}
\label{sec:noise-aware}

The mean Volterra--Riccati policy treats the conditional RFQ forecast curve $m_t$ as a deterministic future intensity. This captures the conditional mean effect of Hawkes excitation but ignores the remaining uncertainty around the future intensity path. When the Hawkes process is persistent, this uncertainty can be economically relevant: two histories with the same current mean forecast may still differ in the conditional dispersion of future RFQ activity. We introduce a perturbative functional delta method correction at the level of the deterministic-intensity Riccati value map. For a deterministic non-negative forecast curve $m$, let $\mathcal V_0(t,q;m)$ denote the value obtained by solving the mean Volterra--Riccati system on $[t,T]$ with forecast curve $m$. Thus the mean approximation in the previous subsection is
\[
v_0(t,q;m_t)=\mathcal V_0(t,q;m_t).
\]
Conditional on $\mathcal F_t$, the forecast curve $m_t$ is deterministic, but the realized future intensity process fluctuates around this conditional mean due to future RFQ arrivals.
Let
\begin{equation}
\xi_t(u)=\lambda_{t+u}-m_t(u), \qquad u\geq 0,
\end{equation}
so that $\mathbb E_t[\xi_t(u)]=0$, and define the conditional covariance kernel
\begin{equation}
\Sigma_t(u,v)=\operatorname{Cov}_t(\lambda_{t+u},\lambda_{t+v}).
\end{equation}
The covariance is understood componentwise over side-size types, i.e.
\begin{equation}
\Sigma_{t,ij}(u,v) = \mathbb E_t\!\left[\xi_{t,i}(u)\xi_{t,j}(v)\right].
\end{equation}
Assume that the map $m\mapsto\mathcal V_0(t,q;m)$ is twice differentiable in the neighbourhood of $m_t$, in the sense of functional derivatives. Denote its first and second sensitivities by
\begin{equation}
D_i(t,q;u) = \left. \frac{\delta \mathcal V_0(t,q;m)}{\delta m_i(u)} \right|_{m=m_t},
\qquad
K_{ij}(t,q;u,v) =
\left. \frac{\delta^2 \mathcal V_0(t,q;m)} {\delta m_i(u)\,\delta m_j(v)} \right|_{m=m_t}.
\end{equation}
A second-order expansion of the deterministic-intensity value map around the conditional mean forecast gives
\begin{equation}
\begin{aligned}
\mathcal V_0(t,q;m_t+\xi_t) \approx\;& v_0(t,q;m_t) + \sum_i\int_0^{T-t} D_i(t,q;u)\xi_{t,i}(u)\,du  \\
&+ \frac12 \sum_{i,j} \int_0^{T-t}\int_0^{T-t} K_{ij}(t,q;u,v)\xi_{t,i}(u)\xi_{t,j}(v)\,du\,dv .
\label{eq:second-order-functional}
\end{aligned}
\end{equation}
Taking the conditional expectation removes the linear term and yields the noise-aware approximation
\begin{equation}
\bar V(t,q) \approx v_0(t,q;m_t) + \frac12 \sum_{i,j} \int_0^{T-t}\int_0^{T-t}
K_{ij}(t,q;u,v)\Sigma_{t,ij}(u,v)\,du\,dv .
\label{eq:jensen-correction}
\end{equation}
This is a local covariance correction to the conditional mean approximation. It is not an exact value function and is not claimed to restore a dynamic-programming or time-consistency property. Its sign is governed by the curvature of the deterministic-intensity value map with respect to the forecast curve; it is not assumed to be positive in general. Its usefulness is assessed against the exact exponential Hawkes benchmark below.

The corresponding deterministic noise-aware shadow price is
\begin{equation}
\bar p_i(t,q) = \frac{\bar V(t,q)-\bar V(t,q+\epsilon_i z_i)}{z_i}, \qquad i=(s,k),
\label{eq:noise-aware-shadow}
\end{equation}
and the quote is computed from the exact Hamiltonian optimizer,
\[
\delta_i(t,q)=\delta_i^*(\bar p_i(t,q)).
\]
Thus the Hamiltonian expansion is still used only to approximate the
continuation value, while the final quote remains generated by the original
win-probability model.

In implementation, the sensitivities $D$ and $K$ can be obtained by differentiating the Riccati system with respect to a finite-dimensional parametrization of the forecast curve. In an exponential Markovian benchmark, the same quantities can equivalently be obtained by differentiating the lifted Riccati coefficients with respect to the Hawkes memory state. The resulting policy is deterministic conditional on the information available at time $t$: it uses both the conditional mean forecast $m_t$ and the conditional covariance kernel $\Sigma_t$, but it does not yet incorporate the discrete post-request forecast update when valuing the conversion of the RFQ currently being answered. The next approximation level introduces this realized state feedback explicitly.

\subsection{State-feedback stochastic Volterra--Riccati approximation}

The strongest approximation retains the Volterra--Riccati continuation value but updates the forecast curve after observed RFQs. In the exact path-dependent HJB, the relevant shadow price after a request of type $i=(s,k)$ is evaluated under the post-request history. The forecast-based analogue is therefore to evaluate the continuation value under the post-request conditional forecast. Let $\mathcal F_t^{i,+}$ denote the information state immediately after observing a type-$i$ RFQ at time $t$, and define
\begin{equation}
m_t^{i,+}(u)=\mathbb E\!\left[\lambda_{t+u}\mid\mathcal F_t^{i,+}\right].
\label{eq:generic-post-request-forecast}
\end{equation}
For the Hawkes model,
\begin{equation}
m_t^{i,+}(u)=m_t(u)+\rho_i(u),
\label{eq:hawkes-post-request-forecast}
\end{equation}
where $\rho_i$ is the forecast response defined in \eqref{eq:forecast-response}. The three approximation levels reveal a useful separation between the request-flow model and the Riccati control layer: the mean approximation requires $m_t$, the covariance correction additionally requires $\Sigma_t$, and state feedback requires the post-request update $m_t^{i,+}$. Hawkes dynamics provide these objects through the Volterra equation and resolvent, but another exogenous-intensity model could in principle supply the same conditional inputs. The mean state-feedback shadow price is then
\begin{equation}
p_i^{\mathrm{SF},0}(t,q) \approx \frac{v_0(t,q;m_t^{i,+})-v_0(t,q+\epsilon_i z_i;m_t^{i,+})}{z_i}.
\label{eq:state-feedback-shadow}
\end{equation}
Equivalently, this is the mean Volterra--Riccati value evaluated not at the pre-request forecast $m_t$, but at the forecast curve that already incorporates the information contained in the newly observed RFQ.

For analytical interpretation, specialize to Hawkes so that $m_t^{i,+}-m_t=\rho_i$, and linearize this expression around the pre-request forecast $m_t$. Using the first functional sensitivity $D$ of the deterministic-intensity Riccati value map, introduced in the previous subsection, gives
\begin{equation}
p_i^{\mathrm{SF},0}(t,q) \approx \Delta_i v_0(t,q;m_t)
+ \frac{1}{z_i} \sum_j \int_0^{T-t} 
\Big[ D_j(t,q;u) - D_j(t,q+\epsilon_i z_i;u) \Big]\rho_{i,j}(u)\,du,
\label{eq:sf-linearized}
\end{equation}
where, for $i=(s,k)$, we write $\Delta_i=\Delta_k^s$, $z_i=z_k$ and
$\epsilon_i=\epsilon_s$. The first term is the mean-forecast shadow price. The second term is the incremental information value of the RFQ: the request is not treated only as a possible inventory jump, but also as a signal that changes the conditional distribution of future RFQ flow.

The same idea can be combined with the noise-aware approximation. Let $\bar V(t,q;m,\Sigma)$ denote the covariance-corrected value obtained from the second-order approximation in the previous subsection when the conditional forecast and covariance are $(m,\Sigma)$. If $\Sigma_t^{i,+}$ denotes the conditional covariance kernel after observing a type-$i$ RFQ, the noise-aware state-feedback shadow price is
\begin{equation}
p_i^{\mathrm{SF}}(t,q) \approx
\frac{\bar V(t,q;m_t^{i,+},\Sigma_t^{i,+}) - \bar V(t,q+\epsilon_i z_i;m_t^{i,+},\Sigma_t^{i,+})}{z_i}.
\end{equation}
For Hawkes, $m_t^{i,+}=m_t+\rho_i$. In applications where the covariance update is ignored or approximated by its pre-request value, one may set $\Sigma_t^{i,+}\approx\Sigma_t$. If this noise-aware expression is linearized, additional derivative terms arise from the covariance-corrected value map and, if retained, from the post-request
covariance update. Thus the displayed linear formula above should be read specifically as the first-order linearization of the mean state-feedback rule.

The quote is finally computed from the exact Hamiltonian optimizer,
\[
\delta_i(t,q)=\delta_i^*(p_i^{\mathrm{SF}}(t,q)),
\]
or from $\delta_i^*(p_i^{\mathrm{SF},0}(t,q))$ in the mean state-feedback version. Hence the Volterra--Riccati approximation affects the continuation-value shadow price, while the final quote remains generated by the original win-probability model.

This state-feedback term is absent from a pure mean-forecast policy. It is the main channel through which the approximation closes the gap to the exact lifted HJB in directional Hawkes regimes. In a general-kernel Hawkes model, the state variable is the updated forecast curve. In an exponential benchmark, the same update is represented by the finite-dimensional Hawkes memory state.

\begin{figure}[t]
\centering
\begin{tikzpicture}[
    node distance=8mm,
    box/.style={draw=blue!55!black, rounded corners=2mm, very thick, align=center, text width=0.74\linewidth, minimum height=11mm, fill=blue!4},
    base/.style={draw=gray!70!black, rounded corners=2mm, thick, align=center, text width=0.56\linewidth, minimum height=8mm, fill=gray!5},
    arrow/.style={-{Latex[length=3mm]}, very thick, draw=blue!55!black},
    dasharrow/.style={-{Latex[length=3mm]}, thick, dashed, draw=gray!70!black}
]
\node[box] (mean) {Mean Volterra--Riccati\\\footnotesize conditional mean forecast $m_t(u)$};
\node[box, below=of mean] (noise) {Noise-aware Volterra--Riccati\\\footnotesize mean forecast plus conditional covariance $\Sigma_t(u,v)$};
\node[box, below=of noise] (state) {State-feedback stochastic Volterra--Riccati\\\footnotesize update the forecast curve after observed RFQs};
\node[base, above=of mean] (pois) {Poisson benchmark\\\footnotesize deterministic baseline intensity, no Hawkes memory};
\node[base, below=of state] (exact) {Exact lifted HJB validation benchmark\\\footnotesize available for exponential kernels; high-dimensional otherwise};
\draw[dasharrow] (pois) -- (mean);
\draw[arrow] (mean) -- (noise);
\draw[arrow] (noise) -- (state);
\draw[dasharrow] (state) -- (exact);
\end{tikzpicture}
\caption{Approximation hierarchy. The modelling problem is Hawkes-driven and generally non-Markovian. The exact lifted HJB is used only as a validation benchmark in kernels where such a lift is finite-dimensional; the Poisson model is a misspecified comparator.}
\label{fig:hierarchy}
\end{figure}
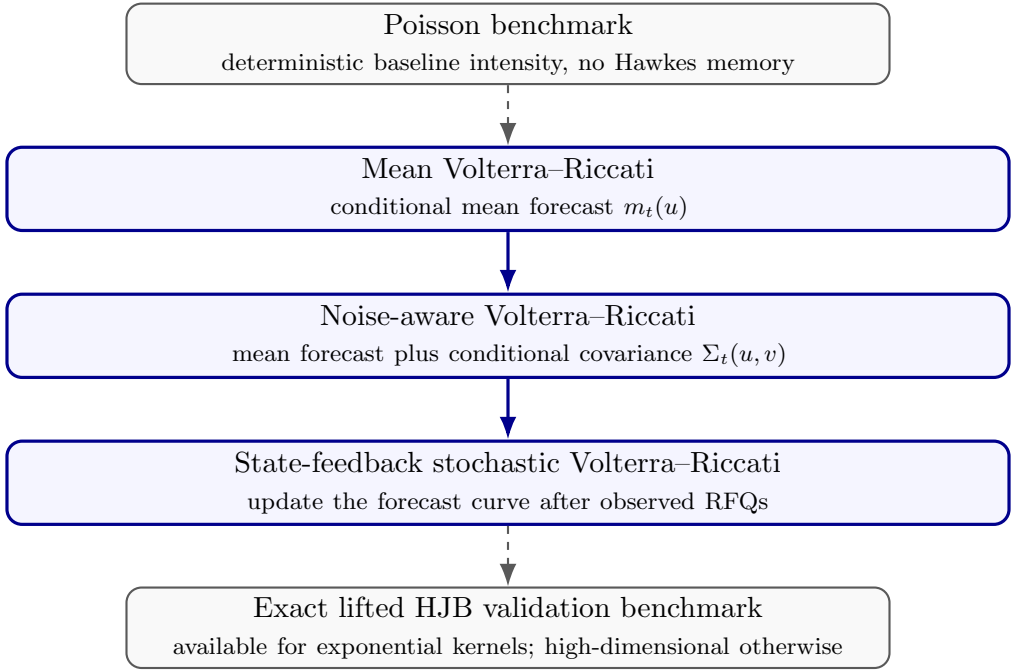

Figure \ref{fig:hierarchy} summarizes the approximation hierarchy.

\section{Exponential Hawkes validation benchmark}

The exponential Hawkes model permits an exact finite-dimensional dynamic-programming benchmark and thus can serve to test the Volterra-Riccati approximation hierarchy in a setting where the true optimal quote can still be computed directly. In this validation experiment we use a unit RFQ size, $z_1 = 1$, and therefore suppress the size index. A fill on side $s \in \{b, a\}$ changes inventory from $q$ to $q + \epsilon_s$. This unit-size benchmark isolates the effect of Hawkes memory and avoids mixing memory effects with size-discretisation effects.

For the lifted exponential benchmark, let $X_t$ denote the scalar Hawkes memory. We write
\begin{equation}
    \dd X_t=-\beta X_t\,\dd t + \sum_{s \in \{b,a\}} \alpha_s\,\dd M_t^s, \qquad \lambda_t^s = \lambda^s(X_t),
    \label{eq:exp-hawkes-memory}
\end{equation}
where $\alpha_s$ is the excitation generated by observing a side-$s$ RFQ. The memory state jumps from $x$ to $x+\alpha_s$ as a result, irrespective of whether the dealer wins the request. The inventory changes only if the request is won. For example, the directional one-sided benchmark corresponds to $\alpha_b = \alpha$ and $\alpha_a = 0$, with $\lambda^b(x) = \mu^b + x$ and $\lambda^a(x) = \mu^a$. A symmetric common-mode benchmark may instead use $\alpha_b = \alpha_a = \alpha$.

Since $(q_t, X_t)$ is Markovian, the exact value function can be written as $V(t, q, x)$. Using the unit-size Hamiltonian $h_1$ defined in \eqref{eq:bg-hamiltonian}, the lifted HJB is
\begin{align}
0 &= \partial_t V(t, q, x) - \beta x\, \partial_x V(t, q, x) - \kappa q^2 \nonumber \\
+ \sum_s \lambda^s(x) \Bigg[ V(t, q, x+\alpha_s) &- V(t, q, x) + h_1 \Big( V(t, q, x+\alpha_s) - V(t, q+\epsilon_s, x+\alpha_s) \Big) \Bigg],
\label{eq:lifted-exp-hjb-unit}
\end{align}
with terminal condition
\begin{equation}
V(T, q, x) = -\ell(q).
\end{equation}
The first difference inside the brackets, $V(t, q, x + \alpha_s) - V(t, q, x)$, is the value effect of observing the RFQ and updating the Hawkes memory even when the dealer does not win the request. The Hamiltonian term is the incremental value of optimally converting that request into a unit fill. Equation \eqref{eq:lifted-exp-hjb-unit} is therefore the exact Markovian control problem for the exponential-memory benchmark. After truncating the inventory and memory domains, it can be solved backward on a grid in $(q, x)$ and compared with the approximation hierarchy.

The conditional mean and variance of the Hawkes memory satisfy closed equations.  In the scalar one-sided case where
\[
\lambda_t = \mu + X_t, \qquad \dd X_t = -\beta X_t\,\dd t + \alpha\, \dd M_t,
\]
one has
\begin{align}
    \dot{x}_t &= \alpha\mu-(\beta-\alpha)x_t,\label{eq:mean-exp}\\
    \dot{y}_t &= -2(\beta-\alpha)y_t+\alpha^2(\mu+x_t),\label{eq:var-exp}
\end{align}
where $x_t=\mathbb{E}[X_t\mid X_0]$ and $y_t=\operatorname{Var}(X_t\mid X_0)$.  The branching ratio is $\eta=\alpha/\beta$.  As $\eta$ approaches one, intensity uncertainty becomes persistent and the mean forecast alone can become insufficient.

Three regimes are used to isolate distinct mechanisms.

\begin{itemize}[leftmargin=2em]
    \item \textbf{Benign common-mode burst.}  A side-symmetric Hawkes burst increases total opportunity flow but does not create strong signed inventory pressure.  The mean forecast is expected to capture most of the value.
    \item \textbf{Near-critical common-mode burst.}  The same common-mode structure is made persistent and noisy by taking a high branching ratio.  This stresses the noise-aware correction.
    \item \textbf{Directional one-sided burst.}  One side of the market is self-exciting, for example
    \[
        \lambda_t^b=\mu^b + X_t,
        \qquad \lambda_t^a = \mu^a,
        \qquad \dd X_t=-\beta X_t\,\dd t+\alpha\,\dd M_t^b.
    \]
    This creates signed inventory pressure.  In this regime the realized memory state is economically important, and the state-feedback stochastic Volterra--Riccati policy is expected to be closest to the exact HJB.
\end{itemize}

The policies compared in each regime are: a (misspecified) Poisson policy using baseline intensities, the mean Volterra--Riccati policy, the mean plus variance correction, the state-feedback stochastic Volterra--Riccati policy, and the exact lifted HJB policy used as the benchmark.

\subsection{Numerical results}

We do not pursue a general well-posedness theorem for the full path-dependent Hawkes control problem. Throughout the numerical experiments, admissible quotes are restricted to a compact interval and the response functions are smooth and bounded by the exogenous Hawkes activity. Hence the controlled intensities and the finite-horizon objective are well defined under the parameter regimes considered. In the exponential memory benchmark, after truncating the inventory and Hawkes-state domains, the exact lifted HJB reduces to a finite system of backward ordinary differential equations. The Volterra--Riccati approximations are likewise obtained from finite-dimensional backward Riccati systems with bounded deterministic coefficients. The focus of the paper is therefore not a general convergence theorem, but a controlled numerical validation of the approximation hierarchy.

\begin{table}[t]
\centering
\caption{Key numerical parameters for the three validation regimes. The common-mode cases use symmetric side intensities, while the directional case excites only one side of the RFQ flow.}
\label{tab:scenario-parameters}
\begin{tabular}{lrrrr}
\toprule
Parameter & Description & Benign & Near-critical & Directional \\
\midrule
$q_0$ & Initial inventory & 10 & 10 & 0 \\
$X_0$ & Hawkes memory & 120 & 120 & 120 \\
$\mu^b$ & Bid RFQ intensity & 10 & 1.5 & 5 \\
$\mu^a$ & Ask RFQ intensity & 10 & 1.5 & 20 \\
$\beta$ & Hawkes decay & 8 & 10 & 10 \\
$\alpha$ & Hawkes excitation & 5.2 & 9.2 & 9.2 \\
$\eta$ & Branching ratio & 0.65 & 0.92 & 0.92 \\
\bottomrule
\end{tabular}
\end{table}

The exact benchmark solves the lifted HJB backward on a grid in $(q,X)$.  The Hamiltonian optimizer is precomputed on a shadow-price grid and then used to recover the optimal quotes. To avoid giving the memory-free model an artificial numerical disadvantage, we do not use the Riccati approximation for the Poisson policy but rather solve the underlying exact one-dimensional HJB on the inventory grid. Thus the Poisson dealer is optimal within their own misspecified memory-free model, but their quotes depend only on time and inventory, not on the realized Hawkes memory. In the unit-size case the Poisson value function $V^P(t, q)$ solves the familiar
\begin{equation}
0 = \partial_t V^P(t, q) - \kappa q^2 + \sum_s \bar{\lambda}^s(t) h_1 \Big( V^P(t, q) - V^P(t, q + \epsilon_s)\Big),
\end{equation}
with terminal condition $V^P(T, q) = - \ell(q)$. Here $\bar{\lambda}^s(t)$ denotes the deterministic intensity used by the Poisson dealer. In the experiments below this is the baseline intensity of the corresponding side, so the Poisson policy deliberately ignores the realized Hawkes excitation. 

The remaining policies are approximate. The mean Volterra--Riccati policy solves a deterministic time-inhomogeneous Riccati system driven by the conditional mean forecast, the noise-aware policy adds the covariance correction, and the state-feedback policy updates the forecast or lifted memory state in the quote rule. Since the Hawkes memory is one-dimensional in this validation experiment, the
conditional forecast curve is represented by the current memory state $X_t$.

\begin{figure}[h]
\centering
\includegraphics[width=0.98\linewidth]{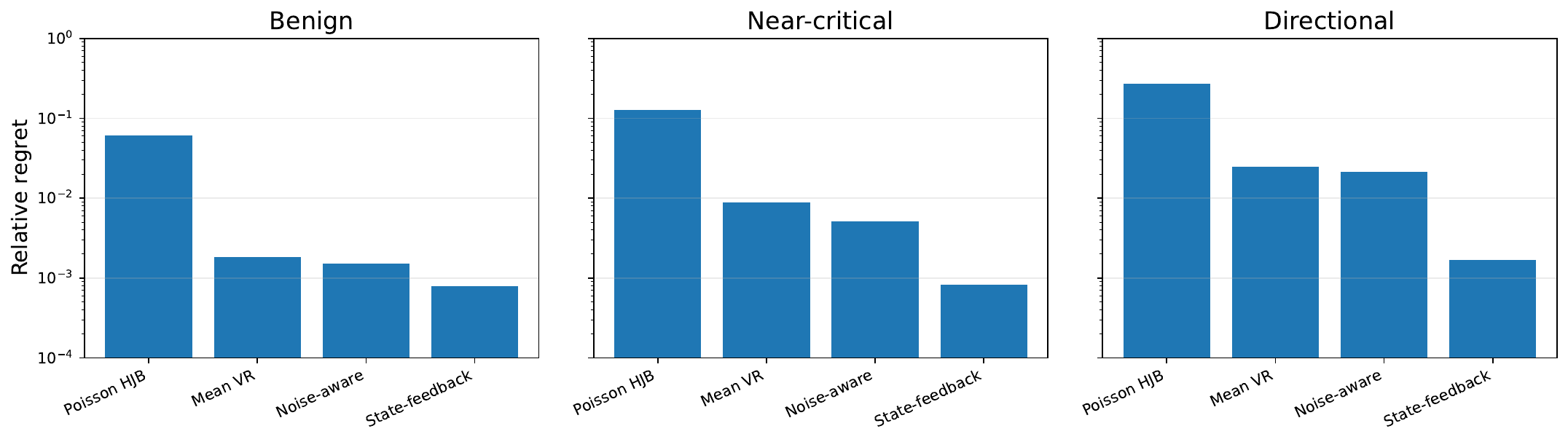}
\caption{Relative paired regret against the exact lifted HJB across the three exponential Hawkes validation regimes.  The Poisson policy is exact within the misspecified memory-free model. The Volterra--Riccati policies progressively add mean forecasting, intensity uncertainty correction and realized memory state feedback. State feedback gives the closest approximation, particularly in the directional regime, where the current Hawkes memory contains economically relevant information about future signed RFQ flow.}
\label{fig:paired-regrets}
\end{figure}

Monte Carlo simulation in this study is used only for out-of-sample evaluation of the already computed controls under the true Hawkes dynamics. All policies are evaluated under the same simulated Hawkes environments.  Common random numbers are used across policies, so accuracy is reported through paired regret, i.e. the difference between the exact lifted HJB objective and the approximate policy objective, where both objectives are evaluated pathwise on the same simulated RFQ and fill randomness and then averaged. This removes most of the sampling noise in comparisons between close policies. Since the objective scale differs across the three scenarios, the relative paired regret (with respect to the exact objective) is also reported.

Table~\ref{tab:scenario-parameters} reports the key parameters used in the three validation regimes. In the numerical benchmarks, one day is the unit of the chosen model clock; RFQ intensities and the volatility-derived penalty are expressed per unit of that same clock. Trade size and inventory units correspond to one million notional and price increments are in basis points. In addition, the period $T = 1$ day was used for all scenarios with timestep $\Delta t = 10^{-4}$. Calculations were performed on an integer inventory grid with $q_{\max} = 50$ million and a 1000-point memory grid with $X_{\max} = 800$. The quote offset grid was $\delta \in [-5, 20]$ bp with 5000 points and shadow price grid was $p \in [-10, 14]$ with 2400 points. Sigmoid win probability function $f(\delta)=\left\{1+\exp\bigl(\delta-1.2)\bigr)\right\}^{-1}$ was assumed throughout. The penalty coefficients were $\kappa =\kappa_T = 0.2$, where we assumed quadratic terminal penalty form $\ell(q) = \kappa_T q^2$. Finally, the number of generated Monte Carlo paths was $2\cdot 10^4$. 

\begin{table}[t]
\centering
\caption{Policy performance in the exponential Hawkes validation benchmark.
Objectives and paired regrets are computed under common simulated Hawkes paths.
Objectives are reported with standard error, absolute regrets with 95\% confidence.}
\label{tab:policy-summary}
\begin{tabular}{ccccc}
\toprule
Scenario & Policy & Objective & Absolute Regret & Relative Regret (\%) \\
\midrule
Benign & Exact lifted HJB & 42.46 $\pm$ 0.12 & 0.000 $\pm$ 0.000 & 0.000 \\
Benign & Poisson HJB & 39.86 $\pm$ 0.11 & 2.606 $\pm$ 0.059 & 6.138 \\
Benign & Mean VR & 42.38 $\pm$ 0.12 & 0.077 $\pm$ 0.027 & 0.182 \\
Benign & Noise-aware VR & 42.40 $\pm$ 0.12 & 0.064 $\pm$ 0.026 & 0.151 \\
Benign & State-feedback VR & 42.43 $\pm$ 0.12 & 0.034 $\pm$ 0.024 & 0.080 \\
\midrule
Near-critical & Exact lifted HJB & 48.80 $\pm$ 0.24 & 0.000 $\pm$ 0.000 & 0.000 \\
Near-critical & Poisson HJB & 42.58 $\pm$ 0.22 & 6.218 $\pm$ 0.085 & 12.742 \\
Near-critical & Mean VR & 48.37 $\pm$ 0.25 & 0.432 $\pm$ 0.042 & 0.884 \\
Near-critical & Noise-aware VR & 48.55 $\pm$ 0.24 & 0.250 $\pm$ 0.036 & 0.513 \\
Near-critical & State-feedback VR & 48.76 $\pm$ 0.24 & 0.041 $\pm$ 0.025 & 0.083 \\
\midrule
Directional & Exact lifted HJB & 46.20 $\pm$ 0.10 & 0.000 $\pm$ 0.000 & 0.000 \\
Directional & Poisson HJB & 33.58 $\pm$ 0.08 & 12.617 $\pm$ 0.119 & 27.312 \\
Directional & Mean VR & 45.04 $\pm$ 0.10 & 1.154 $\pm$ 0.054 & 2.498 \\
Directional & Noise-aware VR & 45.20 $\pm$ 0.10 & 0.997 $\pm$ 0.049 & 2.159 \\
Directional & State-feedback VR & 46.12 $\pm$ 0.10 & 0.079 $\pm$ 0.020 & 0.170 \\
\bottomrule
\end{tabular}
\end{table}

\begin{figure}[h]
\centering
\includegraphics[width=0.62\linewidth]{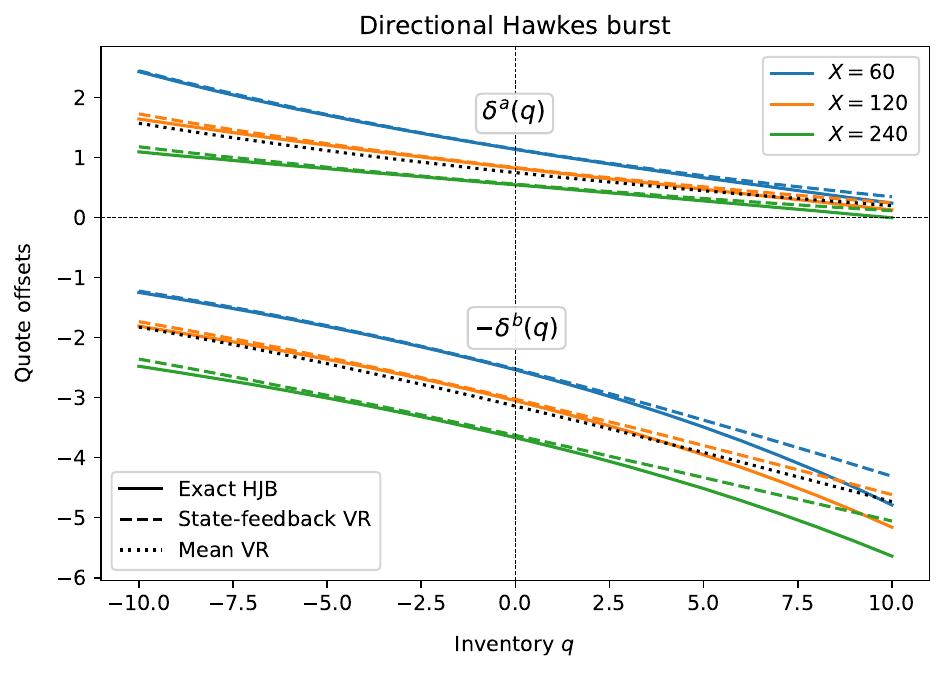}
\caption{Directional Hawkes benchmarks: ask offsets $\delta^a(q)$ and minus bid offsets $-\delta^b(q)$ at the initial time. Coloured curves correspond to realized Hawkes-memory states $X=60, 120, 240$. Solid curves are the exact lifted HJB and dashed curves are the state-feedback stochastic Volterra--Riccati approximation. The dotted black curves show the mean Volterra--Riccati policy, which does not depend on the realized Hawkes state.}
\label{fig:directional-quotes}
\end{figure}

Figure~\ref{fig:paired-regrets} demonstrates the relative paired regret in the three regimes. The Poisson policy is the exact one-dimensional HJB solution for the memory-free model, so its error is due to model misspecification rather than to a weak numerical implementation. In the common-mode regimes, Hawkes excitation mainly changes the amount of future RFQ opportunity flow. The mean Volterra--Riccati policy therefore captures much of the effect, while covariance correction becomes more valuable when the Hawkes process is more persistent and intensity uncertainty is larger. In the directional regime, however, the realized Hawkes memory is itself a control-relevant state variable. The state-feedback Volterra--Riccati policy is then closest to the exact lifted HJB because it updates the quote rule with the realized memory state, rather than using only a deterministic forecast. Table~\ref{tab:policy-summary} reports the complementary absolute quantities and errors. For each policy and regime it shows the Monte Carlo objective, absolute paired regret and relative regret. The objective column gives the scale of the control problem, while the regret column gives the loss in the original objective units.

Figure~\ref{fig:directional-quotes} explains the mechanism behind the directional regime ranking. It shows quote offsets at the initial time as functions of inventory for three realized Hawkes memory states, $X=60, 120, 240$. We plot the ask offset $\delta^a(q)$ together with minus the bid offset $-\delta^b(q)$, so that both sides can be compared on a common vertical scale. The exact lifted HJB and the state-feedback Volterra--Riccati policy both react to the realized Hawkes state: as $X$ increases, the quote skew changes because the dealer anticipates a stronger conditional imbalance in future RFQ arrivals. The mean Volterra--Riccati policy, on the other hand, does not condition on the realized value of $X$. Its bid-ask skew comes from the deterministic directional mean forecast used when the Riccati system is solved, not from the displayed state slices.

The numerical evidence therefore separates three effects. First, solving the Poisson baseline exactly confirms that the large directional-regime loss is not a numerical artifact: it is the cost of ignoring Hawkes memory. Second, deterministic Volterra forecasting captures the average effect of excitation and already improves substantially over the memory-free policy. Third, when the Hawkes state carries directional information, the realized memory must enter the quote rule. The state-feedback Volterra--Riccati approximation captures this last mechanism and closely tracks the exact lifted HJB, while remaining applicable to general kernels for which a low-dimensional lifted HJB is unavailable.

\section{Power-law RFQ memory and endogenous OTC impact}
\label{sec:power-law-impact}

The exponential Hawkes benchmark above validates the approximation hierarchy in a setting where an exact lifted HJB can be computed.  We now use the same state-feedback Volterra--Riccati rule in the intended setting: a long-memory RFQ environment for which no low-dimensional exact dynamic program is available.

The purpose of the experiment is not only numerical.  The Volterra formulation also gives a transparent mechanism for OTC market impact.  A directional RFQ burst changes the conditional forecast of future RFQ flow.  Through the state-feedback shadow price \eqref{eq:state-feedback-shadow}, this forecast change immediately modifies quotes.  Thus quote impact is generated endogenously by optimal conditioning on RFQ information, rather than being imposed as an exogenous price-impact function.

\subsection{Linearized RFQ-information impact}

Consider a side-size type $i=(s,k)$ and a current forecast curve $m_t$.  The state-feedback quote uses
\[
    \delta_i^*(p_i^{\rm SF}(t,q)),
\]
where $p_i^{\rm SF}$ is given by \eqref{eq:state-feedback-shadow}.  Suppose now that, at time zero, the dealer observes a directional RFQ burst of type $i_0$.  At a later time $t$, the additional conditional forecast at future lag $u$ is, to first order,
\[
    \rho_{i_0}(t+u),
\]
where $\rho_{i_0}$ is the Volterra forecast response defined in \eqref{eq:forecast-response}.  Linearising \eqref{eq:state-feedback-shadow} around the unperturbed forecast gives the incremental shadow-price impact
\begin{equation}
\begin{split}
    \Delta p_i(t,q) \approx \frac{1}{z_i}\sum_j\int_0^{T-t} \left[ D_j(t,q;u) - D_j(t,q+\epsilon_i z_i;u) \right] \rho_{i_0,j}(t+u)\,\dd u .
\end{split}
\label{eq:linearized-shadow-impact}
\end{equation}
Here $z_i = z_k$, $\epsilon_i = \epsilon_s$ for $i = (s, k)$.
This is the analytical counterpart of the numerical quote response. The kernel $\rho_{i_0}$ describes the change in future conditional intensities following an observed RFQ; the sensitivity difference in brackets describes how future intensity changes affect the continuation value of filling the current RFQ. The impact is therefore a Volterra filter of the RFQ forecast response, without requiring a causal interpretation of the fitted Hawkes branching representation.

The quote impact follows by differentiating the Hamiltonian optimizer:
\begin{equation}
    \Delta \delta_i(t,q) \approx \left(\delta_i^*\right)'\!\left(p_i^0(t,q)\right) \Delta p_i(t,q),
    \label{eq:linearized-quote-impact}
\end{equation}
where $p_i^0(t,q)$ is the corresponding baseline shadow price.  Equation \eqref{eq:linearized-quote-impact} is only a local formula; in the numerical experiment below the final quote is still computed with the exact optimizer $\delta_i^*(\cdot)$.  The formula is useful because it identifies the source of the impact and its tail behaviour.

To see this tail behaviour, suppose the Hawkes kernel is subcritical and has a power-law tail.  Write the matrix of kernel masses as
\[
    \mathcal B=\int_0^\infty \Phi(u)\,\dd u,
\]
with spectral radius $r(\mathcal B)<1$.
If, for a slowly varying function $L$ and exponent $\chi>1$,
\[
    \Phi(u)\sim C_\Phi u^{-\chi}L(u),
    \qquad u\to\infty,
\]
then the Volterra response $\rho_i$ has the same tail order.  More precisely, the Hawkes resolvent asymptotics give
\begin{equation}
    \rho_i(u) \sim (I-\mathcal B)^{-1} C_\Phi (I-\mathcal B)^{-1} e_i\,  u^{-\chi}L(u), \qquad u\to\infty .
    \label{eq:resolvent-power-law-tail}
\end{equation}
If the Riccati sensitivity kernel in \eqref{eq:linearized-shadow-impact} is integrable in the future lag $u$, then the slowly varying tail can be pulled through the finite-memory Volterra filter. In a stationary setting, or in a finite-horizon computation with a sufficiently long continuation horizon so that the relevant Riccati sensitivities are effectively away from the terminal boundary, the convolution in \eqref{eq:linearized-shadow-impact} preserves the tail order of the resolvent response. Hence, for fixed inventory $q$ and side-size type $i$,
\begin{equation}
    \Delta \delta_i(t,q) \sim  C^{\rm imp}_{i,i_0}(q)\,  t^{-\chi}L(t), \qquad t\to\infty ,
    \label{eq:quote-impact-tail}
\end{equation}
Here $C^{\rm imp}_{i,i_0}(q)$ is the leading-order impact amplitude obtained by combining three terms: the tail constant of the Hawkes resolvent response to the initial burst of type $i_0$, the lag-integrated Riccati sensitivity of the current shadow price for type $i$, and the local derivative of the quote optimizer $(\delta_i^\ast)'$ at the baseline shadow price. Its sign and magnitude depend on the burst direction, the quoted side and size, and the current inventory, but the time decay is inherited from the RFQ-memory kernel.

Thus the Volterra--Riccati quote impact has the same long-lag decay exponent as the conditional RFQ-flow response. Inventory feedback and the nonlinear sigmoid optimizer affect the amplitude and short-time shape, but not the leading power-law tail under the above long-horizon approximation. The general mechanism is forecast-based rather than Hawkes-causal: any alternative request-flow model producing a post-shock conditional forecast with the same long-lag asymptotics would be filtered through the same Riccati sensitivity and would generate the same leading decay mechanism, subject to the corresponding regularity assumptions. The explicit amplitude and resolvent asymptotics in \eqref{eq:resolvent-power-law-tail}, however, are specific to the Hawkes specification.

\subsection{Numerical experiment}

We now illustrate the mechanism in a two-sided marked RFQ market with a
power-law-like Hawkes memory. Consider the RFQ size ladder $\mathcal{Z} = \{1,2,5\}$ with probabilities $(0.65,0.25,0.10)$, so that $\mathbb E[z]=1.65$ and $\mathbb E[z^2]=4.15$. As before, size is in units of million notional. Conditional on an RFQ, the dealer wins with sigmoid probability $f(\delta)=\left\{1+\exp\bigl(3(\delta-1)\bigr)\right\}^{-1}$.
The corresponding myopic zero-shadow-price quote is $\delta^\ast(0)=0.85$ bp, corresponding to $f(\delta^\ast(0))=0.61$.

The side-level Hawkes kernel is represented by a scalar exponential mixture
multiplied by a branching matrix $\mathcal B$,
\[
\Phi(u)=g_N(u)\mathcal B,
\qquad
g_N(u)=\sum_{n=1}^N w_n\beta_n e^{-\beta_n u},
\qquad
\sum_{n=1}^N w_n=1.
\]
With this normalization, $\mathcal B$ is the same kernel-mass matrix as in the asymptotic discussion above. In the experiment
\[
\mathcal B=
\begin{pmatrix}
0.55 & 0.05\\
0.05 & 0.55
\end{pmatrix},
\qquad
r(\mathcal B)=0.60<1.
\]
Thus same-side excitation dominates cross-side excitation, while the Hawkes
system remains subcritical. The background intensities are symmetric,
$\mu^b=\mu^a=155$, giving stationary prospective RFQ intensities
$\bar\lambda^b=\bar\lambda^a=388$ per day. These values correspond to a
target unconditional executed notional of approximately
$2\,\bar\lambda^s f(\delta^\ast(0))\mathbb E[z] = 780$ million per day.

The mixture contains $N=8$ exponential factors. The decay rates $\beta_n$ are
logarithmically spaced on $[2,2500]$, and the weights are chosen as
\[
w_n=
\frac{\beta_n^{-\nu_{\rm mix}}}
{\sum_{l=1}^N\beta_\ell^{-\nu_{\rm mix}}},
\qquad
\nu_{\rm mix}=0.65.
\]
This construction allocates more mass to slower factors and makes the finite
mixture behave like a slowly decaying power law over the relevant intraday
range. Since the grid is logarithmic,
\[
g_N(u)\approx
C\int \beta^{-\nu_{\rm mix}}\beta e^{-\beta u}\,d\log\beta
=
C\int \beta^{-\nu_{\rm mix}} e^{-\beta u}\,d\beta
\propto u^{-(1-\nu_{\rm mix})}
\]
over the interior scaling range. 
Since the mixture is finite, the power-law behaviour should be
read as an intermediate-range approximation: the fastest exponential creates a
short-time cut-off and the slowest exponential eventually dominates at very long
lags.

\begin{figure}[h]
\centering
\includegraphics[width=0.49\textwidth]{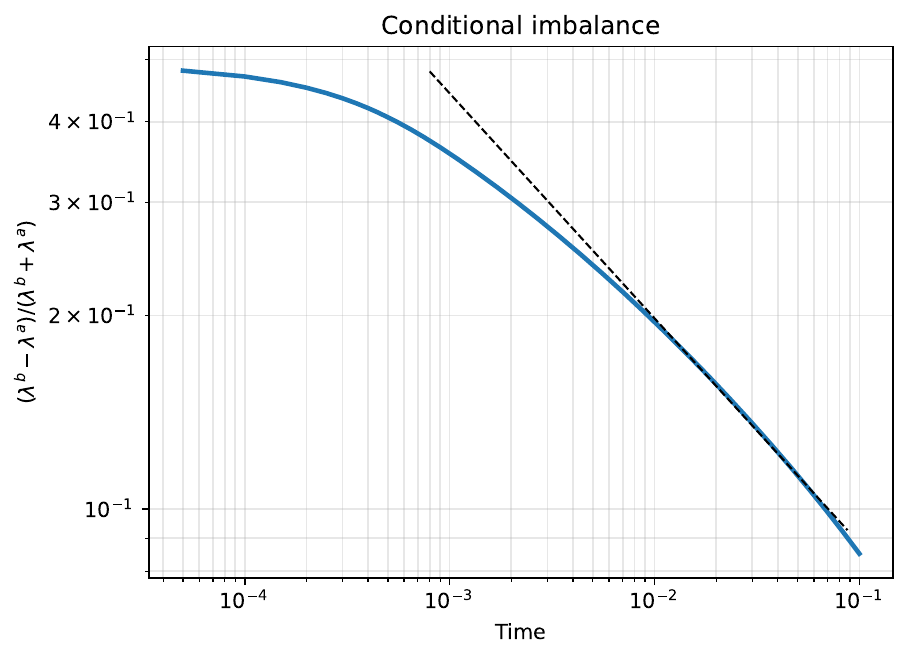}
\includegraphics[width=0.49\textwidth]{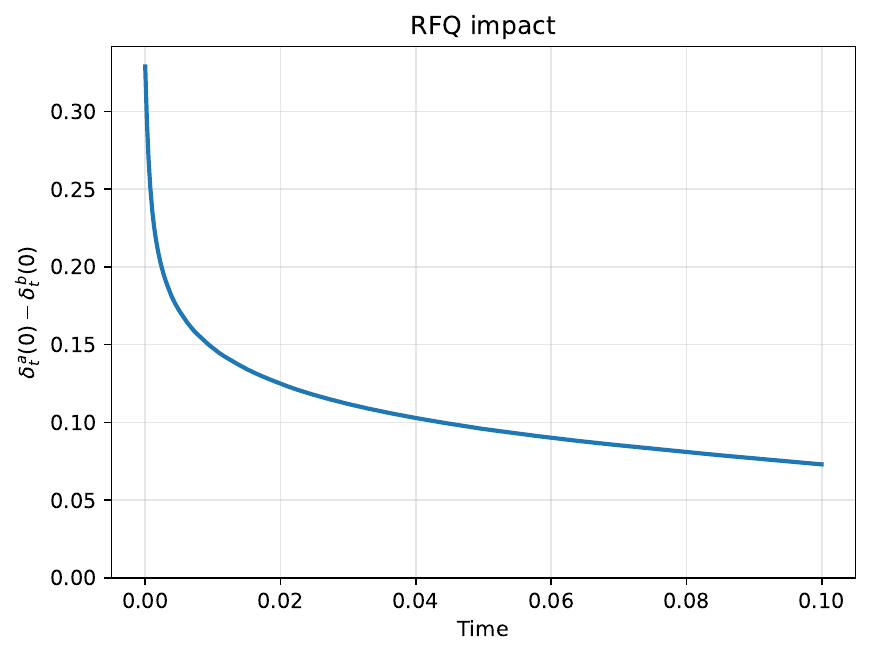}\\
\caption{Power-law RFQ memory (conditional imbalance, left, log-log) and direct quote impact at zero inventory (quote skew, right). The guide slope (dashed) in the left panel is $-(1-\nu_{\rm mix})=-0.35$, as described in the text. A one-sided RFQ burst creates a persistent conditional imbalance in future request flow.  The Poisson policy does not react to this information.  The state-feedback Volterra--Riccati policy converts the forecast imbalance into a persistent quote skew, providing an endogenous source of OTC market impact.}
\label{fig:powerlaw-mechanism}
\end{figure}

The state-feedback Volterra--Riccati dealer uses the conditional forecast curve
generated by this Hawkes model, but we do not solve a lifted HJB in the
exponential factors. The Poisson benchmark uses the same RFQ size distribution,
win probability and inventory penalty, and is calibrated to the same stationary
side intensities $\bar\lambda^b=\bar\lambda^a$. Its misspecification is
therefore not the unconditional level of RFQ activity, but the absence of
conditioning on the realized Hawkes memory and on the resulting signed
imbalance.

The running inventory penalty is set to $\kappa = 2.0$ and the terminal penalty is $\ell(q)=\kappa_T q^2$ with $\kappa_T=0.01$.
The Monte Carlo evaluation horizon is $T_{\rm eval}=0.1$ days, discretised with
$\Delta t=5\cdot 10^{-5}$, giving $2000$ time steps. Before applying the
information shock, the Hawkes process is simulated for a burn-in period
$T_{\rm burn}=0.1$ and the burn-in observations are discarded. This avoids the
artificial empty-history effect that would arise if the process were started
with zero memory at the beginning of the experiment. The quote computation uses an additional continuation horizon $T_{\rm cont}=0.1$. In practice, the backward Volterra--Riccati system is solved on the extended horizon $T_{\rm eval}+T_{\rm cont}$, while only the first $T_{\rm eval}$ units of the resulting policy are used in the reported simulation. This continuation horizon is therefore not an additional evaluation period; it is a numerical device that reduces terminal-horizon distortions in the displayed post-burst window. The conditional forecast curve in the Volterra--Riccati solver is discretised with $N_{\rm f}=12$ future-lag grid points. This forecast grid is distinct from both the Monte Carlo time step
$\Delta t$ and the $N=8$ exponential factors used to represent the Hawkes
kernel. The quote optimizer is tabulated on $\delta\in[-1, 12]$ bp and $p\in[-15, 25]$, using respectively $30001$ quote grid points and $5201$ shadow price grid points.

At time zero, after the burn-in period, the dealer observes a one-sided RFQ
burst. Without loss of generality, we consider a bid-side burst, corresponding
to clients selling to the dealer, so that fills tend to increase dealer
inventory. Default burst size is $50$ RFQs, unless specified otherwise. Directional perturbation changes the signed conditional distribution of future RFQs, not merely the total activity level. This is the cleanest setting in which to isolate information-driven OTC impact. All Monte Carlo comparisons use common
random numbers across the Poisson and state-feedback Volterra--Riccati policies,
with $10^5$ simulated paths.

Figure \ref{fig:powerlaw-mechanism} illustrates the mechanism.  The left panel plots the conditional RFQ imbalance after the burst.  Its slow decay reflects the power-law Volterra response. The right panel plots the corresponding direct RFQ-information quote impact at zero inventory.  The Poisson dealer has no response because their future flow forecast is unchanged by the burst.  The state-feedback Volterra--Riccati dealer immediately skews quotes because the burst changes the continuation-value shadow price.

\begin{figure}[h]
\centering
\includegraphics[width=0.49\textwidth]{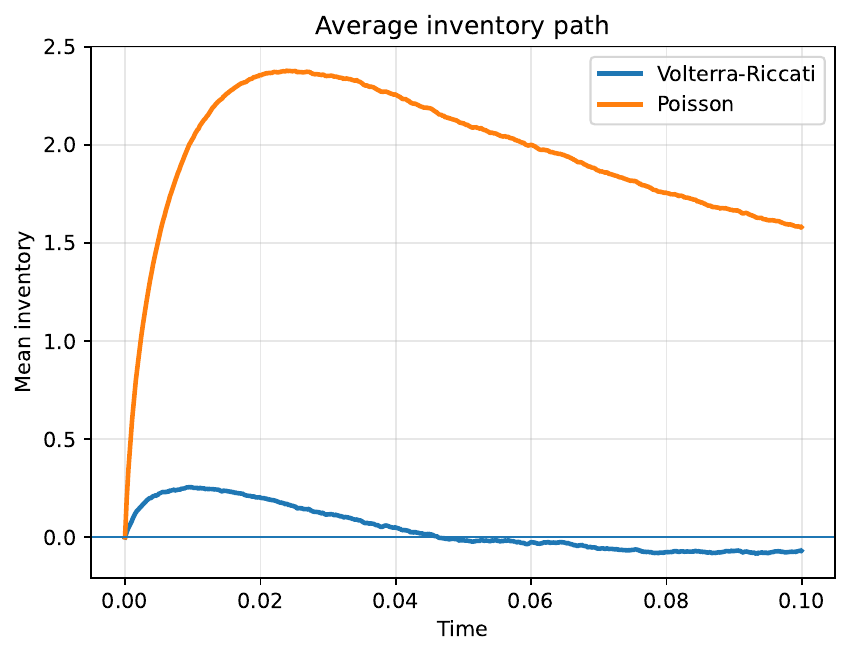}
\includegraphics[width=0.49\textwidth]{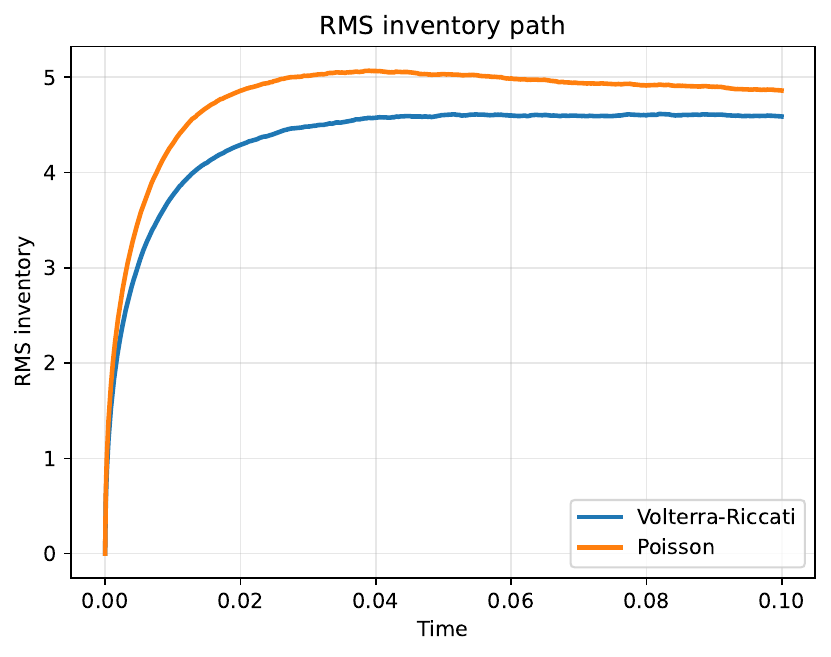}\\
\caption{Inventory response after a directional RFQ burst.  The Poisson dealer ignores the conditional future-flow imbalance and accumulates directional inventory.  The state-feedback Volterra--Riccati dealer uses the updated forecast curve in the continuation-value shadow price and reduces both mean and RMS inventory.}
\label{fig:powerlaw-inventory}
\end{figure}

Figure \ref{fig:powerlaw-inventory} demonstrates the inventory consequence of the same mechanism.  The Poisson dealer is pulled into directional inventory after the burst.  The state-feedback dealer anticipates the conditional future flow and quotes defensively, reducing both the average inventory displacement and the root mean square (RMS) inventory.  This is the most robust economic effect in the experiment: conditioning on the RFQ memory primarily improves inventory-risk control.

\begin{figure}[h]
\centering
\includegraphics[width=0.62\textwidth]{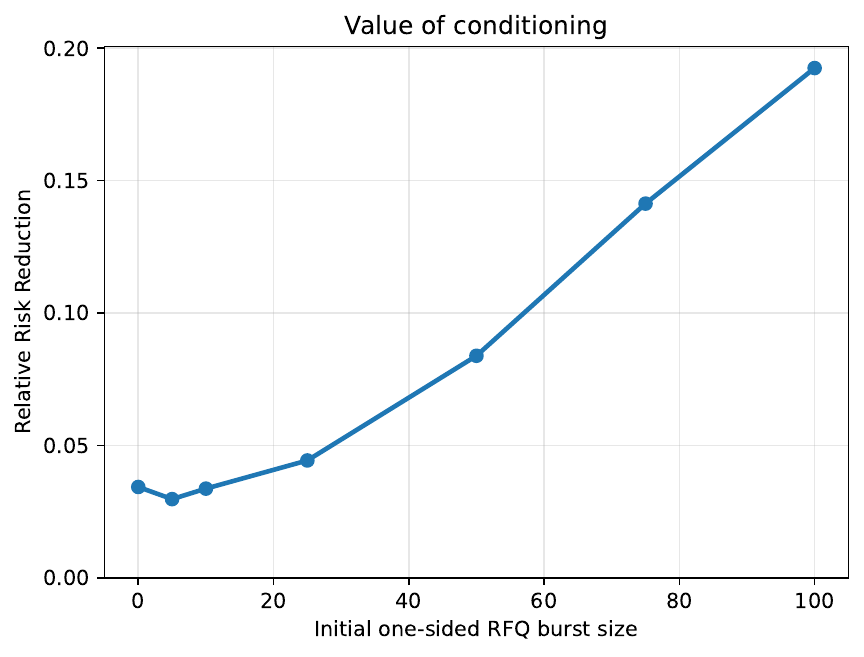}
\caption{Value of conditioning on the RFQ memory.  As the initial directional burst becomes larger, the conditional imbalance in future RFQ flow becomes more informative.  The state-feedback Volterra--Riccati policy uses this information to reduce inventory risk and P\&L volatility relative to the Poisson no-conditioning benchmark.}
\label{fig:powerlaw-value-conditioning}
\end{figure}

Finally, Figure \ref{fig:powerlaw-value-conditioning} varies the size of the initial one-sided burst.  The value of conditioning increases with the strength of the observed RFQ imbalance.  We report relative reduction in P\&L standard deviation, comparing state-feedback Volterra--Riccati with the Poisson benchmark, i.e. 1-StDev(P\&L$_\text{VR}$)/StDev(P\&L$_\text{Poisson}$). Mean P\&L differences are less stable and are not the main message; the robust effect is the reduction of inventory and P\&L risk when the flow history is informative.

These findings are consistent with, and complementary to, earlier academic work on persistent order flow and market impact.  \citet{Jaisson2015Impact} interprets market impact as anticipation of future order-flow imbalance; our mechanism gives an OTC/RFQ version of the same idea, where the relevant information is not an exchange trade print but the dealer's observed request flow.  The power-law decay in Figure \ref{fig:powerlaw-mechanism} is also in line with the no-arbitrage and long-memory impact mechanisms studied by \citet{JusselinRosenbaum2020}, and with the broader Hawkes-based connection between order-flow persistence, impact and volatility developed by \citet{MuhleKarbeChahdiRosenbaumSzymanski2026}.  The OTC interpretation is particularly close in spirit to \citet{EislerBouchaud2016}, who document price impact without a limit order book.  The difference here is that the impact is derived from a dealer's optimal quote response: persistent RFQ memory changes the conditional distribution of future flow, and the Volterra--Riccati shadow price converts this information into a persistent quote skew. The uncovered impact mechanism is different from, and complementary to, the inventory-based impact of a trade \citep{Barzykin2026}.

This experiment should be read together with the exponential Hawkes validation.  In the benchmark case the same approximation hierarchy was compared to the exact lifted HJB, and the state-feedback Volterra--Riccati rule was closest to the exact policy in directional regimes.  In the present power-law case, an exact lifted benchmark is not available at comparable complexity.  The role of the experiment is therefore to apply the validated approximation to the genuinely non-Markovian setting.  The analytical formula \eqref{eq:linearized-shadow-impact} explains why the numerical quote response is persistent, while the Monte Carlo results show that this endogenous OTC impact has economic value through lower inventory and P\&L risk.

\section{Concluding remarks}

This paper develops a Volterra--Riccati approximation for market making when RFQ arrivals are persistent and history-dependent. The main practical message is that, in an OTC market, the relevant state is not only inventory and time. Recent RFQ history changes the conditional distribution of future client demand, and this information should enter the dealer's continuation-value shadow price. In directional regimes, ignoring this state can lead to systematically misplaced quotes and unnecessary inventory risk. The state-feedback Volterra--Riccati rule provides a tractable way to incorporate such information without solving a high-dimensional lifted HJB.

The fitted Hawkes model should be understood as a reduced-form conditional-intensity model rather than a structural account of how client requests cause one another. A common exogenous driver may generate several RFQs and, if it is absent from the baseline intensity, the fitted branching representation may attribute part of this dependence to endogenous excitation. Generalized Langevin models provide a complementary effective-memory viewpoint: persistent observable dynamics can arise after latent degrees of freedom have been reduced out, without requiring literal event-to-event contagion \citep{Itkin2026BeyondRough,Itkin2026LeanMarketron}. A natural empirical extension is therefore to compare pure Hawkes dynamics with latent-factor or Cox specifications, and potentially retain Hawkes excitation only for residual event-to-event dependence after persistent common drivers have been modelled.

Several further practical qualifications are important. The empirical motivation in this paper uses two-way RFQ activity, whereas the most economically valuable state variable for market making is directional flow pressure. In real RFQ markets, directionality is not always directly observed. Some RFQs are one-way, and in other cases dealers can infer likely direction from previous client activity, historical conversion patterns, skewed response behaviour, or contemporaneous public market activity. Nevertheless, directionality assessment is a statistical filtering problem, not an assumption that can be taken for granted. A production implementation would therefore require a joint filtering layer that maps RFQ messages, client history, quote outcomes and public market variables into a conditional forecast of future signed RFQ flow.

A second limitation is the separation between request arrivals and win probabilities. The model assumes that RFQs follow the exogenous conditional intensity model, while the dealer's win probability conditional on an RFQ remains governed by the quoted offset and is not itself directly affected by the Hawkes memory. This is a useful baseline because it separates information arrival from execution conversion, and the same conceptual decomposition applies to request-for-stream protocols. It is not, however, the only possible mechanism. In stressed or one-sided markets, RFQ bursts may coincide with changes in competition, client urgency, quote toxicity or last-look behaviour, all of which can alter the quote-response function. Extending the framework to state-dependent win probabilities is therefore a natural next step.

The mean and covariance-corrected Volterra--Riccati constructions are approximations rather than exact value functions. A frozen conditional forecast does not satisfy an intertemporal dynamic-programming principle; the implemented policy obtains feedback by reconditioning and resolving as the history evolves. The exponential benchmark provides numerical evidence that this receding horizon construction is accurate in the regimes studied, but a general error or convergence theory remains open. One possible route to a more systematic hierarchy is through expansions of point-process functionals, including the pseudo-chaotic expansions for counting processes developed by \citet{HillairetReveillac2024}; pursuing that connection is beyond the scope of the present paper.

A third extension concerns the effect of fills and hedging. In the baseline model, requests drive Hawkes memory and fills are controlled thinnings of those requests. This prevents the dealer from mechanically controlling future RFQ intensity through their own executions. In practice, however, executed trades may contain additional information: fills reveal stronger client intent than unfilled RFQs, may update relationship-level beliefs, and may be followed by externalisation trades that interact with public market activity. Similarly, RFQ intensity may be co-excited with price moves, volatility, market depth or observable order-flow imbalance in lit venues. A richer model would therefore treat RFQs, fills, hedging trades and public market signals as a coupled marked point process or Volterra system.

These limitations indicate where the approximation is most useful. Exact dynamic programming for such enriched non-Markovian models is unlikely to be feasible. The value of the Volterra--Riccati approach is that it preserves an interpretable conditional forecast state and converts it into quote adjustments through a tractable backward Riccati calculation. The Riccati layer is partly modular with respect to the request-flow model: Hawkes dynamics provide the required conditional objects explicitly, while other predictive flow models could in principle feed the same control calculation. This makes the construction a practical bridge between persistent RFQ flow and implementable market-making controls while leaving open the structural origin of the persistence. Future work should focus on latent-factor and directional RFQ filtering, separating common persistent drivers from residual excitation, state-dependent response functions, joint RFQ--price excitation, and empirical validation on realized quoting and fill data.

\section*{Acknowledgment}
The author is grateful to Richard Anthony (HSBC) for support throughout the project, and to Olivier Guéant (Université Paris Cité), Andrey Itkin (NYU Tandon School of Engineering) and Eyal Neuman (Imperial College London) for fruitful discussions and valuable comments. The views expressed are those of the author and do not necessarily reflect the views or practices at HSBC.

\end{document}